\documentclass[preprint,showpacs,preprintnumbers,nofootinbib,amsmath,amssymb,aps,prc,longbibliography]{revtex4-1}
\usepackage{graphicx}% Include figure files
\usepackage{dcolumn}% Align table columns on decimal points
\usepackage{bm}% bold math
\begin{document}
\title{Giant dipole resonance in highly excited nuclei}
\thanks{Talk at the Int'l Nucl. Phys. Conf. INPC 2013, Florence, Italy, 2-7 June 2013.}
\author{Nguyen Dinh Dang}
  \email{dang@riken.jp}
 \affiliation{Theoretical Nuclear Physics Laboratory, RIKEN Nishina Center
for Accelerator-Based Science,
2-1 Hirosawa, Wako City, 351-0198 Saitama, Japan\\
and Institute for Nuclear Science and Technique, Hanoi, Vietnam\\
}
\date{\today}% It is always \today, today,
\begin{abstract}
The evolution of the giant dipole resonance's (GDRÕs) width and shape at finite temperature $T$ and angular momentum $J$ is described within the framework of the phonon damping model (PDM). The PDM description is compared with the established experimental systematics obtained from heavy-ion fusion and inelastic scattering of light particles on heavy target nuclei, as well as with predictions by other theoretical approaches. Extended to include the effect of angular momentum $J$, its strength functions have been averaged over the probability distributions of $T$ and $J$ for the heavy-ion fusion-evaporation reaction, which forms the compound nucleus $^{88}$Mo at high $T$ and $J$. The results of theoretical predictions are found in excellent agreement with the experimental data. The predictions by PDM and the heavy-ion fusion data are also employed to predict the viscosity of hot medium and heavy nuclei. 
\end{abstract}
\maketitle
\section{Introduction}
\label{intro}
The GDR built on highly excited compound (CN) nuclei was first observed in 1981~\cite{Newton}, and at present rich experimental systematics has been established for the GDR widths at finite temperature $T$ and angular momentum $J$ in various medium and heavy nuclei formed in heavy ion fusions, deep inelastic scattering of light particles on heavy targets, and $\alpha$ induced fusions\cite{Schiller,Kolkata}. The common features of the hot GDR are: (1) Its energy is nearly independent of $T$ and $J$, (2) Its full width at half maximum (FWHM) remains mostly unchanged in the region of $T\leq$ 1 MeV, but increases sharply with $T$ within 1$\leq T\leq$ 2.5 - 3 MeV, and seems to saturate at $T\geq$ 4 MeV. As a function of $J$, a significant increase in the GDR width is seen only at $J\geq$ 25 - 27$\hbar$. In Ref. \cite{Kusnezov}, by adding the pre-equlibrium $\gamma$ emission in reanalyzing some GDR data, it was claimed that the GDR width does not saturate. However, it was realized later that the pre-equilibrium emission is proportional to the asymmetry between projectiles and targets and lowers the CN excitation energy. This may alter the conclusion on the role of pre-equlibrium emission. The recent measurements in $^{88}$Mo at $T\geq 3$ MeV and $J>$ 40$\hbar$ did not show any significant effect of pre-equilibrium emission on the GDR width~\cite{Maj}.  The evaporation width due to the quantal mechanical uncertainty in the energies of the CN states was also proposed to be added into the total GDR width~\cite{Chomaz}, whose effect may become noticeable only at much higher values of $T$ ($\gg$ 3.3 MeV) and $J$ ($\gg$ 30$\hbar$)~\cite{Gervais}. From the classical representation of the GDR as a damped spring mass system, it is clear that the damping width of the oscillator should be smaller than its frequency otherwise the spring mass system cannot make any oscillation. This means that the GDR width in the classical picture is upper-bounded by its energy. This implies the saturation of the GDR width.

The present contribution summarizes the achievements of the Phonon Damping Model (PDM)~\cite{PDM} in the description of the the GDR width and shape at finite $T$ and $J$. The GDR parameters predicted by the PDM and experimentally extracted are also used to calculate the shear viscosity of finite hot nuclei. 
\\
\\
\\
\section{Damping of GDR in highly excited nuclei}
\label{TJ}
The PDM's Hamiltonian consists of the 
independent single-particle (quasiparticle) field, GDR phonon field, and the coupling between 
them.   The Woods-Saxon potentials at $T =$ 0 are  
used to obtain the single-particle energies $\epsilon_k$. The GDR width $\Gamma(T)$ is a sum: $\Gamma(T)=\Gamma_{\rm Q}+\Gamma_{\rm T}$ of
the quantal width, $\Gamma_{\rm Q}$, and thermal width, $\Gamma_{\rm T}$.
In the presence of superfluid pairing, the quantal and thermal widths 
are given as~\cite{PDM} $\Gamma_{\rm Q}=2\gamma_Q(E_{GDR})=2\pi F_{1}^{2}\sum_{ph}[u_{ph}^{(+)}]^{2}(1-n_{p}-n_{h})
\delta[E_{\rm GDR}-E_{p}-E_{h}]~$,
and
$\Gamma_{\rm T}=2\gamma_T(E_{GDR})=2\pi F_{2}^{2}\sum_{s>s'}[v_{ss'}^{(-)}]^{2}(n_{s'}-n_{s})
\delta[E_{\rm GDR}-E_{s}+E_{s'}]~,$ 
where $u_{ph}^{(+)} = u_pv_h+u_hv_p$, $v_{ss'}^{(-)}=u_su_{s'}-v_sv_{s'}$ ($ss' = pp', hh'$) with
$u_k$ and $v_k$ being the coefficients of Bogolyubov's transformation, $E_k\equiv\sqrt{(\epsilon_k-\lambda)^2+\Delta^2}$, with superfluid pairing gap $\Delta$,
are quasiparticle energies, $n_k$ are quasiparticle occupations numbers, which, for medium and heavy nuclei, can be well approximated with the Fermi-Dirac distribution for
independent quasiparticles, $n_k = [\exp(E_k/T)+1]^{-1}$. The parameter $F_1$ is chosen so that $\Gamma_Q$ at $T=$ 0 is equal to GDR's width at $T=$ 0, whereas the parameter $F_2$ is chosen so that, with varying $T$,  the GDR energy $E_{GDR}$ does not change significantly. The latter is found as the solution of the equation $E_{GDR} - \omega_{q}-P_q(E_{GDR})=0$, where $\omega_q$ is the energy of the GDR phonon before the coupling between the phonon and single-particle mean fields is switched on, and $P_q(\omega)$ is the polarization operator owing to this coupling, whose explicit expression in given in Refs. \cite{PDM}. The GDR strength function is calculated as $S_{q}(\omega) = (1/\pi)[\gamma_Q(\omega) + \gamma_T(\omega)]/\{(\omega-E_{GDR})^2+[\gamma_Q(\omega) + \gamma_T(\omega)]^{2}\}~.$
In numerical calculations the representation $\delta(x) =\lim_{\varepsilon\rightarrow 0}\varepsilon/[\pi(x^{2}+\varepsilon^2)]$ is used for the $\delta$-function with $\epsilon=$ 0.5 MeV.

\begin{figure}
\center{
\includegraphics[width=17cm]{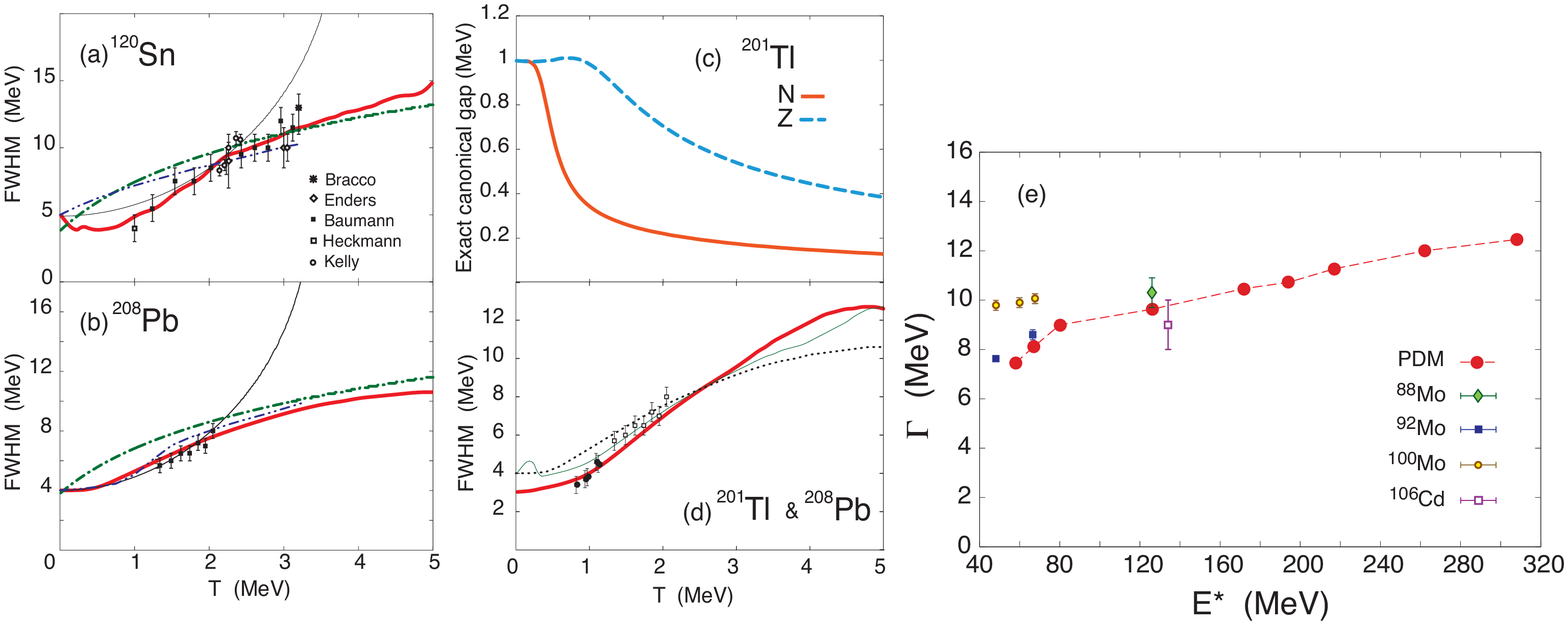}}
\caption{GDR widths for $^{120}$Sn (a) and $^{208}$Pb (b) predicted by the PDM (thick solid), pTSFM (dot-dashed), AM (double dot-dashed), and FLDM (thin solid) as functions of $T$ in comparison with experimental data in tin and lead regions. (c): Exact canonical neutron (N) and proton (Z) pairing gaps for $^{201}$Tl as functions of $T$. (d): GDR width for $^{201}$Tl obtained within the PDM as a function of $T$ (thick solid) including the exact canonical gaps in (c) in comparison with the experimental data for $^{201}$Tl (black circles) and $^{208}$Pb (open boxes). The thin solid line is the PDM result without the effect of thermal pairing. The dotted line is the PDM result for $^{208}$Pb [the same as the thick solid line in (b)]. (e): 
GDR width in the fusion-evaporation reaction $^{48}$Ti + $^{48}$Ca $\rightarrow ^{88}$Mo$^{*}$ as a function of $E^{*}$.  The (red) full circles are PDM predictions, connected with the dashed line to guide the eye. 
\label{width}}
\end{figure}
The GDR widths predicted by the PDM, the two versions of thermal shape fluctuation model (TSFM), namely the phenomenological TSFM (pTSFM) and the adiabatic model (AM), and the Fermi liquid drop model (FLDM) for $^{120}$Sn and $^{208}$Pb are shown in Figs. \ref{width} (a) and \ref{width} (b) in comparison with the experimental systematics. The PDM results for $^{120}$Sn include the effect of non-vanishing thermal pairing gap because of thermal fluctuations owing to finiteness of nuclei. Among the models under consideration, the PDM is the only one that is able to describe well the experimental data in the entire temperature region including $T\leq$ 1 MeV. It is also able to reproduce the very recent data for the GDR width in $^{201}$Tl at 0.8 $\leq T <$ 1.2 MeV [Fig. \ref{width} (d)] after including the exact canonical gaps for neutrons and protons shown in Fig. \ref{width} (c)~\cite{Tl201}.

%%%%%%%%%%%%%%%%%%%%%%%%%%%%%%%%
To describe the non-collective rotation of a spherical nucleus, the $z$-projection $M$ of the total angular momentum $J$ is added into the PDM Hamiltonian as  $- \gamma\hat{M}$, where $\gamma$ is the rotation frequency~\cite{PDMJ}. The latter and the chemical potential are defined, in the absence of pairing,  from the equation $M = \sum_k m_k(f_{k}^{+}-f_{k}^{-})~,$ and  $N =\sum_k(f_{k}^{+}+f_{k}^{-})~$, where $N$ is the particle number and $f_k^{\pm}$ are the single-particle occupation numbers, $f_{k}^{\pm} =1/[\exp(\beta E_k^{\mp})+1]$, and $E_k^{\mp} = \epsilon_k-\lambda\mp\gamma m_k~$. 
 
 The GDR width obtained within the PDM for $^{88}$Mo is plotted against $E^{*}$ in Fig. \ref{width} (e) in comparison with the available GDR experimental widths for molybdenum isotopes. These values are obtained by averaging the GDR strength functions over the probability distributions of $T$ and $J$ for the heavy-ion fusion-evaporation reaction, which forms the compound nucleus $^{88}$Mo at high $T$ and $J$. At $E^{*}\leq$ 80 MeV the increase in the width is rather strong, but at $E^{*}>$ 80 MeV the width increase is weaker because of the saturation of $J_{max}$~\cite{DangCieMaj}. 
%%%%%%%%%%%%%%%%%%%%%%%%%%%%%%%%%%%%%%%%%%%%%%%%
\section{Shear viscosity of hot nuclei}
\label{visco}
In the verification of the condition for applying hydrodynamics to 
nuclear system, the quantum mechanical 
uncertainty principle requires a finite viscosity for any thermal 
fluid. Kovtun, Son and Starinets (KSS)~\cite{KSS} conjectured
that the ratio $\eta/s$ of shear viscosity $\eta$ to the entropy 
volume density $s$ is bounded below for all fluids, namely the value
${\eta}/{s}= {\hbar}/(4\pi k_{B})$  
is the universal lower bound (KSS bound or unit).
From the viewpoint of collective theories, one of the fundamental 
explanations for the giant resonance damping is the friction term (or viscosity) 
of the neutron and proton fluids.  By using the Green-Kubo's relation, it has been shown in Ref. \cite{visco} that the shear viscosity $\eta(T)$ 
at finite $T$ is expressed in terms of the GDR's parameters at zero and finite $T$ as
\begin{equation}
\eta(T)=\eta(0)\frac{\Gamma(T)}{\Gamma(0)}
\frac{E_{GDR}(0)^{2}+[\Gamma(0)/2]^{2}}{E_{GDR}(T)^{2}+[\Gamma(T)/2]^{2}}~.
\label{eta1}
\end{equation}
%%%%%%%%%%%%%%%%%%%%%%%%%%%%%%%%%%%%%%
\begin{figure}
     \center{\includegraphics[width=17.0cm]{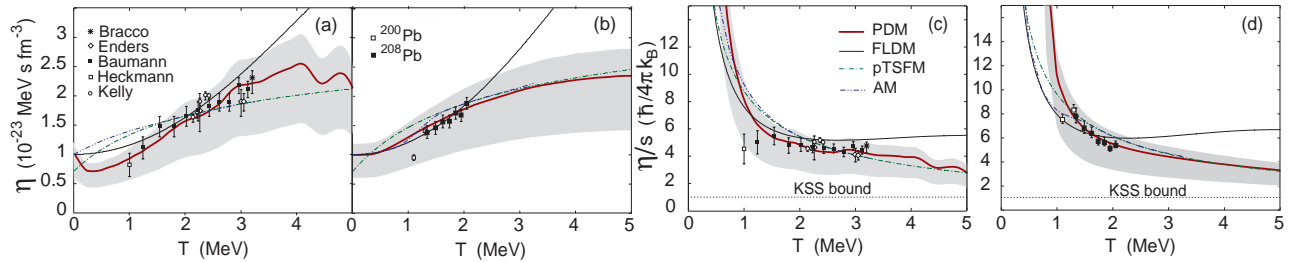}
     \caption{Shear viscosity $\eta(T)$ [(a) and (b)] and ratio 
     $\eta/s$ [(c) and (d)] as functions of $T$ for nuclei in tin [(a) and (c)], and lead [(b) and (d)] regions. The  
     gray areas are the PDM predictions by using $0.6u\leq\eta(0)\leq 
     1.2u$ with $u=$ 10$^{-23}$ Mev s fm$^{-3}$.}
     \label{eta&ratio}}
\end{figure}
%%%%%%%%%%%%%%%%%%%%%%%%%%%%%%%%%%%%%%
The predictions for the shear viscosity $\eta$ and the ratio $\eta/s$ by the PDM,  pTSFM, AM, and 
 FLDM for $^{120}$Sn and $^{208}$Pb are plotted as functions of $T$ 
in Fig. \ref{eta&ratio} in comparison with the empirical results. 
The latter are extracted from the 
experimental systematics for GDR in tin and lead regions~
\cite{Schiller} making use of Eq. (\ref{eta1}).  It is seen in Fig. \ref{eta&ratio} that the predictions by the PDM have the best overall 
agreement with the empirical results. The ratio $\eta/s$ decreases sharply with 
increasing $T$ up to $T\sim$ 1.5 MeV, starting from which the decrease 
gradually slows down to reach (2 - 3) KSS units 
at $T=$ 5 MeV. The FLDM has a similar trend as that of the PDM up to 
$T\sim$ 2 - 3 MeV, but at higher $T$ ($T>$ 3 MeV for $^{120}$Sn or 2 MeV for 
$^{208}$Pb) it produces an increase of both $\eta$ and $\eta/s$ with $T$. 
At $T=$ 5 MeV the FLDM model predicts the ratio $\eta/s$ within (3.7 - 6.5) KSS units, which are 
roughly 1.5 -- 2 times larger than the PDM predictions. The AM and pTSFM show a similar trend for $\eta$ and $\eta/s$. 
However, in order to obtain such similarity, $\eta(0)$ in the pTSFM 
calculations has to be reduced to 0.72$u$ instead of 1$u$. They all 
overestimate $\eta$ at $T<$ 1.5 MeV. Based on these results and on a model-independent estimation, one can conclude that
$\eta/s$ for medium and heavy nuclei at $T=$ 5 MeV is in 
between (1.3 - 4.0) KSS units, which is about (3 - 5) times smaller 
(and of much less uncertainty) that the value between (4 - 19) KSS units predicted by 
the FLDM for heavy nuclei~\cite{Auerbach}, where the same lower value $\eta(0)=$0.6$u$ was used.

\section{Conclusions}
The PDM generates the damping of GDR through its couplings to $ph$ configurations, causing the quantal width, as well as to $pp$ and/or $hh$ configurations, causing the thermal width. This leads to an overall increase in the GDR width at low and moderate $T$, and its saturation at high $T$. At very low T $<$ 1 MeV the GDR width remains nearly constant because of thermal pairing. The PDM predictions agree well with the experimental systematics for the GDR width and shape in various medium and heavy nuclei. The PDM also predicts the shear viscosity to the entropy-density ratio $\eta/s$ between (1.3 - 4.0) KSS units for medium and heavy nuclei at $T=$ 5 MeV, almost the same at that of the quark-gluon-plasma like matter at $T>$ 170 MeV (1.5 - 2.5 KSS) discovered at RHIC and LHC.


\begin{thebibliography}{99}
\bibitem{Newton}J.O. Newton {\it et al.}, Phys. Rev. Lett. {\bf 46}, 1380 (1981).
\bibitem{Schiller}A. Schiller and M. Thoennessen, Atomic Data and 
    Nuclear Data Tables {\bf 93}, 548 (2007). 
  \bibitem{Kolkata}S. Mukhopadhyay {\it et al.}, Phys. Lett. B {\bf 709}, 9 (2012).
 \bibitem{Kusnezov}D. Kusnezov, Y. Alhassid, and K.A. Snover, Phys. 
    Rev. Lett. {\bf 81}, 542 (1998).
\bibitem{Maj}M. Ciemala, Ph.D. thesis (in Polish), Niewodniczanski Institute of Nuclear Physics PAN, Krakow, Report No. 2062/PL (2013) (http://www.ifj.edu.pl/publ/reports/2013/2062.pdf).
\bibitem{Chomaz}Ph. Chomaz, Phys. Lett. B {\bf 347},
1 (1995).
    \bibitem{Gervais}G. Gervais, M. Thoennessen, and W. E. Ormand, Phys. Rev. C
{\bf 58}, R1377 (1998).  
    \bibitem{PDM}N. Dinh Dang and A. Arima, Phys. Rev. Lett {\bf 80}, 4145
    (1998); N. Dinh Dang and A. Arima, Phys. Rev. C {\bf 68}, 044303
    (2003).
    \bibitem{Tl201}N. Dinh Dang and N. Quang Hung, Phys. Rev. C {\bf 86}, 044333 (2012).
    \bibitem{PDMJ}N. Dinh Dang, Phys. Rev. C {\bf 85}, 064323 (2012).
    \bibitem{DangCieMaj}N. Dinh Dang, M. Ciemala, M. Kmiecik, and A. Maj, Phys. Rev. C {\bf  87}, 054313 (2013).
    \bibitem{KSS}P.K. Kovtun, D.T. Son, and A.O. Starinets, Phys. Rev. Lett.
    {\bf 94}, 111601 (2005).
 \bibitem{visco}N. Dinh Dang, Phys. Rev. C {\bf 84}, 034309 (2011).   
	 \bibitem{Auerbach}N. Auerbach and S. Shlomo, Phys. Rev. Lett. {\bf
    103}, 172501 (2009).
\end{thebibliography}
\end{document}